\documentclass[aps,prab,reprint,floatfix,superscriptaddress,amsmath,amssymb]{revtex4-2}

\usepackage{graphicx}
\usepackage{bm}
\usepackage{booktabs}
\usepackage{xcolor}
\usepackage[colorlinks=true,linkcolor=blue,citecolor=blue,urlcolor=blue]{hyperref}

\newcommand{\dd}{\mathrm{d}}

\begin{document}

\title{First-principles simulation of free-electron lasers \\ with the Galilean electromagnetic particle-in-cell method:\\ even harmonics of planar undulators}

\author{Alexander Pukhov}
\affiliation{Institut f\"ur Theoretische Physik I, Heinrich-Heine-Universit\"at D\"usseldorf, 40225 D\"usseldorf, Germany}

\date{\today}

\begin{abstract}
Free-electron laser (FEL) codes average the electron motion over the undulator period and add the radiation harmonics through on-axis coupling factors. This is accurate for the fundamental, but the even harmonics of a planar undulator, which vanish on the axis, are not modelled at all. We show that the Galilean electromagnetic particle-in-cell method (GEM-PIC), which resolves the electron motion within the undulator period and the carrier of the forward radiation field without a Courant limit, simulates FELs on two CPU cores at two to three times the cost of a time-dependent run of the averaged code GENESIS. The quasi-cylindrical code GEM-PIC-RZ reproduces the coherent spontaneous emission of a planar undulator in each azimuthal mode $m = 0$--3 to within 2\,\% of the radiation integral, and seeded amplifiers to within a few per cent of GENESIS run with the same physics. For a seeded planar FEL at 10.6\,nm it predicts 12\,MW of second-harmonic power at saturation, $1.1\times10^{-3}$ of the fundamental; this harmonic is absent from the averaged model. The unaveraged code Puffin, which has no space charge, gives a comparable level, 9 to 11\,MW for the space-charge-widened and the initial beam size, and the same transverse structure.
\end{abstract}

\maketitle

\section{Introduction}\label{sec:intro}

High-gain free-electron lasers (FELs) produce coherent radiation from the terahertz range to hard X-rays~\cite{Pellegrini2016,KimHuangLindberg}. Their design relies on simulation codes such as GENESIS~1.3~\cite{Reiche1999,Genesis4}, which solve the coupled electron and radiation equations after two reductions: the electron motion is averaged over the undulator period, and the radiation is described by a slowly varying envelope at the resonant wavelength $\lambda_r$ and its harmonics. The averaging is the standard and well-tested model of the FEL interaction, valid when the gain per period is small, $\rho \ll 1$, with $\rho$ the Pierce parameter~\cite{Bonifacio1984}. Harmonics enter through coupling factors computed on the undulator axis, the Bessel-function factors $[JJ]_h$ of a planar undulator.

Some physics lies outside this model. In a planar undulator the odd harmonics radiate on the axis, the even harmonics do not: their coupling to an on-axis electron vanishes, and they are driven by the transverse gradient of the bunched beam. Averaged codes therefore omit them; GENESIS~4, for example, skips the source term of even harmonics altogether. Nonlinear harmonic generation in high-gain FELs, in which the fundamental bunching drives harmonic bunching and radiation, has been analysed theoretically~\cite{Huang2000,Freund2000,Geloni2005} and observed experimentally, the second harmonic included~\cite{Tremaine2002,Biedron2002}. Its sensitivity to the beam quality, emittance and size, was studied with codes that add the harmonics to an averaged fundamental~\cite{Biedron2002b}.

Codes that keep the electron motion within the period exist. Puffin solves the three-dimensional FEL equations without period averaging and without the slowly varying envelope approximation~\cite{Campbell2012}; full electromagnetic PIC codes such as MITHRA~\cite{Fallahi2018} solve Maxwell's equations directly, often in a Lorentz-boosted frame, in which the undulator period and the radiation wavelength become comparable~\cite{Vay2007,FawleyVay2009,Fallahi2020}. Full PIC is bound by the Courant condition on the radiation wavelength, and in a boosted frame the beam's own fields and its space charge need care~\cite{FawleyVay2009}.

The Galilean electromagnetic particle-in-cell method (GEM-PIC)~\cite{GEMPIC} removes this limit. It writes Maxwell's equations in the Galilean variables $\zeta = z - ct$ and a slow coordinate $s$, keeps the slow derivative of the forward field, which therefore resolves its carrier, and needs no Courant condition because the field update is semi-implicit. GEM-PIC is not a quasi-static approximation: the forward radiation propagates, diffracts and is amplified with its carrier resolved. It was developed for laser-plasma accelerators~\cite{GEMPIC,GEMPICRZ_paper}; the same equations describe a beam that radiates in an undulator. The step in $s$ resolves the undulator period, the $\zeta$ grid the radiation wavelength. There is no period averaging, no envelope of the forward field and no harmonic model: the electrons follow their orbits in the undulator, and the forward radiation and the fields of the beam, including its space charge, are computed from the same field equations. The scope of the model is set by its treatment of the backward field (Sec.~\ref{sec:disc}): radiation travelling backwards, as in an oscillator, is not described. In this paper, ``unaveraged'' refers to the electron motion, ``carrier-resolved'' to the forward field, and ``first-principles'' to the combination of the two within this field model.

In this paper we use the quasi-cylindrical code GEM-PIC-RZ~\cite{GEMPICRZ_paper,GEMPICRZ} to establish this approach for FELs. Section~\ref{sec:model} summarizes the model and the FEL-specific numerics. Section~\ref{sec:modes} shows that the harmonics of a planar undulator separate by azimuthal mode, which makes them natural to compute in a mode code. Section~\ref{sec:valid} validates the code: the coherent spontaneous emission against the radiation integral, mode by mode, and seeded and self-amplified FELs against GENESIS and scaling laws. Section~\ref{sec:harm} presents a seeded planar FEL with its second and third harmonics and checks the second harmonic with Puffin, Sec.~\ref{sec:averaging} asks when period averaging itself becomes inaccurate, and Sec.~\ref{sec:disc} discusses cost and limits.

\section{Model}\label{sec:model}

The equations and their discretization are given in Refs.~\cite{GEMPIC,GEMPICRZ_paper}; here we summarize what matters for FELs.

\emph{Fields.} GEM-PIC-RZ expands the fields in azimuthal modes, $F = F_0 + 2\,\mathrm{Re}\sum_{m=1}^{M} F_m e^{im\varphi}$, and combines the transverse fields into the forward components $R = E_\perp \pm B_\perp$ and the backward components $L$. The forward field obeys $\partial_s R = \ldots - j_\perp$ with the slow coordinate $s = z$ (GT2); it carries the FEL radiation, which stays at fixed $\zeta$ while the electrons slip back by $\lambda_r$ per period. $L$, $E_z$ and $B_z$ follow from their $\zeta$-equations at each level~\cite{GEMPICRZ_paper}. A level $s$ is the plane $z = s$ of the laboratory, so the beam particles at one level all sit at the same undulator phase.

\emph{Electrons.} Beam particles are advanced in $s$ with the Vay pusher~\cite{Vay2008} in the self-fields and the undulator field, $B_y = B_0 g(z)\cos k_u z$ (planar) or $(B_x, B_y) = B_0 g(z)(\cos k_u z, \sin k_u z)$ (helical), with linear entrance and exit ramps $g(z)$ of one period. Two corrections make the coupling exact to second order in the step $\Delta z$~\cite{GEMPICRZ}: the undulator kick of a step is multiplied by $\mathrm{sinc}(k_u\Delta z/2)$, the exact average of the field over the step, so that the orbit amplitude $K/\gamma$ and therefore the resonance are independent of $\Delta z$; and the transverse current is deposited with the velocity at the time of the position, completing the half step of the leapfrog with the undulator field. Without them the gain length depended on $\Delta z$ to first order: in the helical 10\,nm amplifier of Sec.~\ref{sec:helical} it went from 9.5\,cm without the corrections to 8.7\,cm with the first and 8.3\,cm with both, after which halving $\Delta z$ or $\Delta\zeta$ changed it by less than 1\,\%. A third correction concerns the beam's own field. The beam is advanced from $s^0$ to $s^1$ before the level $s^1$ is solved; pushed in the fields of $s^0$ alone, the push is of first order in the step. In an undulator the large self-field of an intense beam turns with the wiggle angle $(K/\gamma)\cos k_u z$, and the first-order error breaks the cancellation $\bm E + \bm v\times\bm B \approx 0$ that keeps a relativistic beam together. GEM-PIC-RZ therefore pushes the beam a second time, after the level $s^1$ has been solved, from $s^0$ in the mean of the fields of $s^0$ and $s^1$, and deposits its current of $s^1$ anew; the level is not solved again, since the change of the current is of second order. Section~\ref{sec:harm} shows what this changes.

\emph{Self-fields and noise.} A bunch starts with its co-moving field for $\beta < 1$, so the longitudinal and transverse space charge are present from the start. Shot noise is loaded with the beamlet method of Fawley~\cite{Fawley2002}: beamlets of $N_b$ macroparticles at equally spaced phases are quiet for the harmonics $h < N_b$, and their displacements reproduce the bunching $\langle|b_h|^2\rangle = 1/N_e$ of $N_e$ random electrons. The modes $m \ge 1$ vanish on the axis, while the macroparticle noise is largest in the small cells there; the beam is therefore loaded with more macroparticles of smaller charge near the axis: the centres (beamlets) at $r < 4\,\mu$m are sampled 16 times as often with 1/16 of the charge, which halves the noise of the $m = 2$ fields in the first cells for 2.2 times the macroparticles~\cite{GEMPICRZ}.

\emph{Diagnostics.} The radiated power at harmonic $h$ is measured by demodulating each mode of $R$ and $L$ at $\pm h k_r$ over slices of a whole number of $\lambda_r$ (a fraction of a wavelength would leave a spurious bunching of the order of the inverse number of wavelengths); the Poynting flux of mode $m \ge 1$ is $\pi\int(|R_m|^2 - |L_m|^2)\,r\,\dd r$ (summed over the $r$ and $\varphi$ components), half of that for $m = 0$. The bunching $|\langle e^{ihk_r\zeta}\rangle|$ is computed per slice. Demodulation does not remove the field that the bunching at $hk_r$ carries along with the beam; for the even harmonics this field is in $m = 0$, and Sec.~\ref{sec:modes} shows how the radiation is separated from it.

\section{Harmonics and azimuthal modes}\label{sec:modes}

In a planar undulator, $x(t + T_u/2) = -x(t)$ and $y = 0$ along the orbit, with $T_u$ the period of the motion. Rotating the transverse plane by $\pi$ maps the orbit onto itself shifted by half a period; the radiation at harmonic $h$ acquires the factor $(-1)^h$ from the time shift. In cylindrical components the rotation only shifts $\varphi$ by $\pi$, so it multiplies a field of azimuthal mode $m$ by $(-1)^m$. Hence, for a planar field and a beam distribution that is symmetric under this rotation (here: round and centred),
\begin{equation}
\text{odd harmonics} \to \text{odd } m, \qquad \text{even harmonics} \to \text{even } m ,
\end{equation}
i.e.\ a harmonic can only populate modes of its own parity; which of these modes carry power depends on the field structure.
The fundamental and the odd harmonics have an on-axis component in $m = 1$; the even harmonics radiate in $m = 0, 2, \ldots$ and vanish on the axis. For the lowest even-harmonic mode, $E_x \propto x\,f(r)$, the decomposition is $E_r = \tfrac12 r f(1 + \cos 2\varphi)$, $E_\varphi = -\tfrac12 r f \sin 2\varphi$: equal power in $m = 0$ and $m = 2$. A mode code with $M = 2$ thus contains the second harmonic completely, and the relation $P_{m=0} = P_{m=2}$ separates its radiation from the co-moving field of the $2k_r$ bunching, which also lives in $m = 0$. For a beam of finite length, with entrance and exit ramps, the separation holds per frequency band rather than exactly: a small even-mode part appears near $k_r$ (Fig.~\ref{fig:emission}).

\section{Validation}\label{sec:valid}

\subsection{Coherent spontaneous emission at all harmonics}\label{sec:emission}

\emph{Why.} The radiation of a short bunch is the linear response of the field solver to a known current. Its comparison with the far-field radiation integral checks the coupling, the propagation and the mode structure of all harmonics at once, independently of any FEL code.

\emph{Setup.} A planar undulator of 10 periods ($\lambda_u = 10.67$\,mm, $K = 1$) and a bunch of 1\,fC at $\gamma = 100$ ($\lambda_r = 0.8\,\mu$m), with $\sigma_z = 20$\,nm and $\sigma_r = 5\,\mu$m, radiate coherently up to the third harmonic and to angles of $\sim 1.5/\gamma$. The reference is the radiation integral~\cite{Jackson} for the same trajectory (same field, ramps and no natural focusing) times the form factor $\exp(-k^2\sigma_z^2 - k^2\theta^2\sigma_r^2)$, decomposed into azimuthal modes of the far field. GEM-PIC-RZ uses $M = 3$, $\Delta\zeta = \lambda_r/64$, $\Delta z = \lambda_u/40$ and a radial grid from 0.5 to 25\,$\mu$m.

\begin{figure}
\includegraphics[width=\columnwidth]{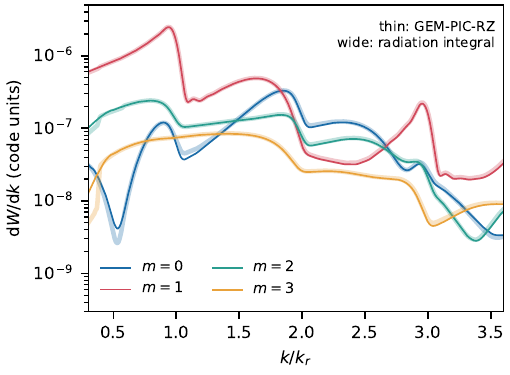}
\caption{Coherent spontaneous emission of a short bunch in a planar undulator (10 periods, $K = 1$, $\gamma = 100$): spectral energy per azimuthal mode, GEM-PIC-RZ (thin) and radiation integral (wide). Odd harmonics in odd modes, even harmonics in even modes.}
\label{fig:emission}
\end{figure}

\emph{Results.} Figure~\ref{fig:emission} and Table~\ref{tab:emission} show the spectra and energies per mode. The energies of the modes $m = 0$--3 differ from the radiation integral by 0.3--1.8\,\%; the radiation integral has a further $9\times10^{-8}$ (4\,\% of the total) in the modes $m \ge 4$, which the run with $M = 3$ does not contain. Changing $\Delta z$, $\Delta r$ or $\Delta\zeta$ by a factor of two changes the mode energies by less than 0.5\,\% (convergence differences, not an uncertainty estimate). One condition matters: radiation emitted at angle $\theta$ advances in $\zeta$ by $\theta^2(L - z)/2$ over the rest of the undulator, 11\,$\mu$m at $\theta = 1.5/\gamma$ here, and the window must leave room for it ahead of the bunch. With only 1.3\,$\mu$m, modes 2 and 3 came out 12--14\,\% low.

\begin{table}
\caption{Coherent emission energy per azimuthal mode (code units, $1 = 2.94\times10^{-7}$\,J). The radiation integral has $9\times10^{-8}$ more in modes $m \ge 4$.}
\label{tab:emission}
\begin{ruledtabular}
\begin{tabular}{lccc}
mode & radiation integral & GEM-PIC-RZ & difference \\
\hline
$m = 0$ & $2.84\times10^{-7}$ & $2.79\times10^{-7}$ & $-1.8$\,\% \\
$m = 1$ & $1.361\times10^{-6}$ & $1.353\times10^{-6}$ & $-0.7$\,\% \\
$m = 2$ & $3.20\times10^{-7}$ & $3.19\times10^{-7}$ & $-0.3$\,\% \\
$m = 3$ & $1.346\times10^{-7}$ & $1.339\times10^{-7}$ & $-0.5$\,\% \\
\end{tabular}
\end{ruledtabular}
\end{table}

\subsection{High-gain amplifiers}\label{sec:helical}

\emph{Seeded, helical.} Two seeded amplifiers in helical undulators, at 10\,nm ($\gamma = 1000$, $\lambda_u = 1$\,cm, $K = 1$, 4\,kA, $\rho = 9.1\times10^{-3}$) and at 0.8\,nm ($\gamma = 5000$, $\lambda_u = 2$\,cm, $\rho = 2.9\times10^{-3}$), were compared with steady-state GENESIS~4~\cite{GEMPICRZ}. The power gain lengths are 8.33 and 38.7\,cm, against 7.92 and 38.4\,cm from GENESIS with the same physics: longitudinal space charge and the widening of the beam by its transverse space charge, which plain GENESIS omits (7.45 and 37.4\,cm). The input decks and the comparison are part of the code repository~\cite{GEMPICRZ}.

\emph{SASE.} A helical SASE FEL ($\gamma = 1000$, 4\,kA, $\sigma_r = 10\,\mu$m, zero emittance, $\lambda_u = 1$\,cm, $K = 1$, $\lambda_r = 10$\,nm, $\rho = 9\times10^{-3}$) with an 8\,$\mu$m flat-top bunch, 14 times $2\pi L_c = \lambda_r/2\rho$, in a 180-period undulator was started once from the shot noise of the real electrons (Fawley beamlets) and once from randomly loaded macroparticles, whose noise power is larger by $N_e/N_\mathrm{mac} = 541$. A uniform focusing channel (2.4\,T/m, from a scan that keeps $\sigma_r$ within 1\,\% over 0.5\,m) balances the transverse space charge; without it the beam widens to 25\,$\mu$m by the end of the undulator. The focused beam keeps its size up to $z \approx 0.6$\,m and then grows slowly, to 11\,$\mu$m at 0.9\,m and 14\,$\mu$m at 1.4\,m, because the nonlinear space charge of the Gaussian profile raises its normalized emittance to 0.1\,$\mu$m by $z = 0.6$\,m (0.12--0.13\,$\mu$m from 0.9\,m on), which the linear channel does not hold; the geometric emittance, $1.2\times10^{-4}\,\mu$m, stays far below $\lambda_r/4\pi = 8\times10^{-4}\,\mu$m; $\rho$ falls to $8.4\times10^{-3}$ at 0.9\,m. We therefore compare the local power gain length, $(\dd\ln P/\dd z)^{-1}$ of the mean power over the flat top, with the Ming Xie fit~\cite{Xie1995} for the local rms size and energy spread (6.4\,cm for the initial beam): once the growth is exponential they agree within 1--17\,\% (ratio 0.99--1.17 at $z = 0.4$--0.8\,m with random loading and 1.00--1.17 at 0.8--1.1\,m with shot noise). In the linear stage the randomly loaded beam has 440--500 times the power of the shot-noise run at the same $z$, close to the ratio 541 of the start-up noise. The randomly loaded beam saturates at $z = 1.15$\,m with 10\,GW. This is 0.53 of the Ming Xie estimate for the initial beam, which the falling local $\rho$ accounts for only in part. Started from the real shot noise, the FEL saturates at $z = 1.62$\,m with 8.3\,GW (229\,$\mu$J), 47\,cm later than with random loading and 23\,\% beyond the Ming Xie saturation length for the initial beam (1.32\,m); by then the beam has grown to 15\,$\mu$m and the local $\rho$ has fallen to $6.9\times10^{-3}$, so the saturation power is 0.45 of the Ming Xie value for the initial beam. At saturation the radiation consists of 11 spikes over the bunch (bunch length$/2\pi L_c = 14$), with a spectral width (FWHM) of $0.8\rho$; the randomly loaded beam, whose start-up noise is 541 times larger and has a shorter exponential stage, gives $1.7\rho$.

\section{Harmonics of a seeded planar FEL}\label{sec:harm}

\emph{Why.} This is the configuration in which an averaged code is incomplete by construction: the second harmonic of a planar undulator is not modelled.

\emph{Setup.} A planar undulator ($\lambda_u = 1$\,cm, $K = 1.5$, 100 periods, no natural focusing in either code) and a cold beam of 4\,kA at $\gamma = 1000$ with $\sigma_r = 10\,\mu$m give $\lambda_r = 10.625$\,nm and $\rho = 8.5\times10^{-3}$. A seed of 10\,MW ($w_0 = 30\,\mu$m) is polarized along the wiggle. GEM-PIC-RZ: $M = 2$, $\Delta\zeta = \lambda_r/32$ over a window of 6.1\,$\mu$m (18\,372 slices), and a uniform radial grid of 1\,$\mu$m up to the conducting wall at $r = 1$\,mm (1000 cells, absorber off). The grid is uniform because a stretched grid reflects transverse wavenumbers $k_\perp \gtrsim 1/\Delta r$ (Sec.~\ref{sec:harm}); the beam is loaded at $x_0 = +x_w = K/(\gamma k_u) = 2.4\,\mu$m, so that its planar orbit after the one-period entrance ramp is centred on the axis, which the parity rule of Sec.~\ref{sec:modes} assumes. The bunch is a flat top of 1.9\,$\mu$m (179\,$\lambda_r$) with $\sin^2$ edges of 0.1\,$\mu$m, longer than the slippage of 1.06\,$\mu$m. It is loaded with 640\,000 macroparticles in Fawley beamlets of 8 (groups of 8 macroparticles at equally spaced phases over one $\lambda_r$, quiet up to the seventh harmonic) plus the refinement near the axis of Sec.~\ref{sec:model} (beamlet centres at $r < 4\,\mu$m sampled 16 times as often with 1/16 of the charge), 1.4 million macroparticles in all. The reference run (``$\lambda_u/32$'') uses $\Delta z = \lambda_u/32$ with the mean-field push of the beam and took 9.1\,h on a 64-core node (44\,GB, dominated by the 1000 radial cells); a run with $\Delta z = \lambda_u/16$ is compared below. The radiation in the slices that met only the flat top is in the steady state and is compared with steady-state GENESIS~4 with fields at $h = 1, 2, 3$.

\emph{Power diagnostics.} The power at harmonic $h$ is demodulated at $\pm hk_r$ per azimuthal mode over slices of $20\lambda_r$ (Sec.~\ref{sec:model}) and averaged over the three slices in the steady-state part of the bunch ($\zeta = 2.76$--3.40\,$\mu$m, between which $P_2$ differs by about 5\,\% rms near saturation); the quoted values are this average at the step of the largest fundamental power (saturation). The power gain length is the inverse slope of $\ln P_1$ between 0.1 and 1\,GW.

\begin{figure}
\includegraphics[width=\columnwidth]{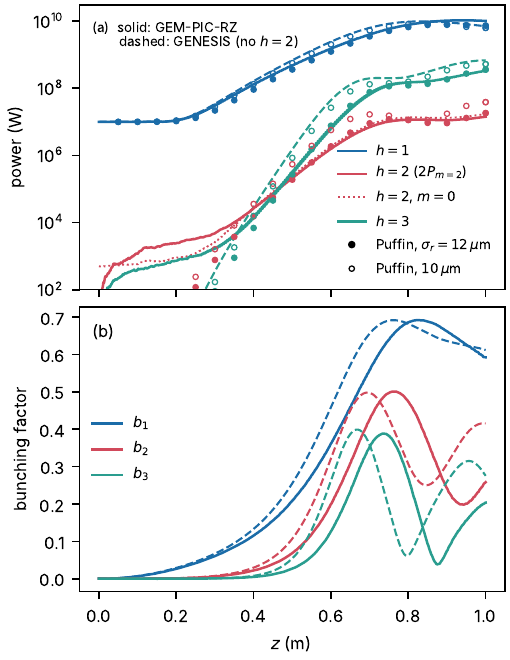}
\caption{Seeded planar FEL at 10.6\,nm: (a) power at the harmonics $h = 1, 2, 3$ and (b) bunching factors, GEM-PIC-RZ (solid, $\Delta z = \lambda_u/32$, mean-field push, refined near the axis) and GENESIS~4 (dashed), in the steady-state part of the bunch, and Puffin (markers; filled: beam of 12\,$\mu$m, the size the GEM-PIC-RZ beam reaches by its space charge; open: 10\,$\mu$m, as at the entrance). GENESIS has no second-harmonic radiation. The second harmonic of GEM-PIC-RZ is $2P_{m=2}$; the excess in $m = 0$ (dotted) is the field of the $2k_r$ bunching that moves with the beam. Before $z \approx 0.35$\,m the second harmonic of GEM-PIC-RZ is spontaneous emission from the shot noise of the beam (Fawley loading), which Puffin, run without shot noise, does not have.}
\label{fig:fel}
\end{figure}

\emph{Results.} Figure~\ref{fig:fel} and Table~\ref{tab:fel}. The fundamental saturates at 10.6\,GW (GENESIS 9.8\,GW); the gain length, 7.9\,cm, is 12\,\% longer than in plain GENESIS, as in the helical case at 10\,nm, where it came from the space charge. The third harmonic reaches 2.3\,\% of the fundamental at saturation (GENESIS 2.8\,\%). With $M = 2$ this is the on-axis ($m = 1$) part of the third harmonic only, with 10.7 points per third-harmonic wavelength. On the earlier, stretched grid (where $P_3/P_1 = 2.24$\,\%), a run with $M = 3$ and $\Delta\zeta = \lambda_r/48$ gives $P_3/P_1 = 2.38$\,\%: the off-axis part in $m = 3$ is 0.2\,\% of $P_3$, and the finer $\zeta$ grid raises $P_3$ by 6\,\%; $P_1$, $P_2$ and $L_g$ change by less than 2\,\%. The remaining difference to GENESIS (2.8\,\%) is of the size of the beam-size effect seen with Puffin below (2.7\,\% and 2.3\,\% for 10 and 12\,$\mu$m).

Both codes bunch the beam at the second harmonic in the same way, with $b_2$ up to 0.5 before saturation. Only GEM-PIC-RZ radiates it. The second harmonic grows from the nonlinear bunching with a gain length of 4.9\,cm, about $L_g/1.6$, and reaches 12\,MW (11.6\,MW) at saturation, $P_2/P_1 = 1.1\times10^{-3}$. \emph{Extraction of $P_2$.} The radiated second harmonic is the $E_x \propto x$ mode (Sec.~\ref{sec:modes}), with equal power in $m = 0$ and $m = 2$. Demodulation keeps, besides the radiation, the field that the $2k_r$ bunching carries along with the beam, and for a round beam this field is in $m = 0$. We therefore use $P_2 = 2P_{m=2}$. Three observations support this. (i) Before the bunching grows ($z < 0.3$\,m, spontaneous emission) $P_{m=0}$ and $P_{m=2}$ agree to within the slice noise (10--20\,\%). (ii) During the exponential growth (0.4--0.7\,m) the excess $P_{m=0} - P_{m=2}$ is proportional to $b_2^2$, constant to $\pm 7$\,\%, as expected for the field of the bunching and not for radiation; near saturation it is 1.4 times $P_{m=2}$. (iii) The co-moving field could leak into $m = 2$ only through the wiggle displacement of the beam, $x_w = K/(\gamma k_u) = 2.4\,\mu$m, which gives an $m = 2$ amplitude of order $(x_w/2\sigma_r)^2 \approx 1.4$\,\% of the $m = 0$ one, i.e.\ $\sim 2\times10^{-4}$ of the excess power: negligible. Puffin, which has no space charge and hence no co-moving field, finds the same equal split of its radiated second harmonic (below). Finally, the saturated $P_2$ does not depend on the early shot-noise emission: in a comparison made on the earlier, stretched grid with the offset orbit (below), a run with a quiet start (beamlets without random displacements) gives 11.2\,MW instead of 11.0\,MW, although its second harmonic is up to 30\,\% lower early on (9\,\% at $z = 0.4$\,m). After saturation the excess is no longer proportional to $b_2^2$, and $P_2$ there is less certain. A small $m = 1$ part (0.3\,\% of the second harmonic) changes by a factor of three within each period: it is the field driven by the wiggling transverse current at $2k_r$, not radiation.

\emph{Independent check with Puffin.} The unaveraged code Puffin~\cite{Campbell2012} radiates the even harmonics as well. We ran it on the same case in its periodic (steady-state) mode, with a slice of $2\lambda_r$, 48 nodes per $\lambda_r$, a transverse mesh of $161^2$ nodes over $300\,\mu$m, 32 steps per period, a quiet beam of 256\,000 macroparticles, the same seed and no natural focusing. Puffin has no space charge. Its gain length, 6.9\,cm, agrees with plain GENESIS (7.0\,cm), as expected without space charge. Its second harmonic is purely $x$-polarized, odd in $x$, and splits equally between $m = 0$ and $m = 2$ (to 1\,\%), as the parity argument of Sec.~\ref{sec:modes} predicts; at saturation it is $P_2 = 11$\,MW, $P_2/P_1 = 1.3\times10^{-3}$, against 12\,MW and $1.1\times10^{-3}$ from GEM-PIC-RZ (Table~\ref{tab:fel}). The third harmonic is 2.7\,\% of the fundamental. The beam of GEM-PIC-RZ widens by its space charge from 10 to 12\,$\mu$m before saturation; in Puffin it keeps its size. Puffin with a 12\,$\mu$m beam (same current) gives $L_g = 7.7$\,cm, $P_2 = 9.1$\,MW, $P_2/P_1 = 1.15\times10^{-3}$ and $P_3/P_1 = 2.3\,\%$. GEM-PIC-RZ lies 6\,\% above the Puffin run with the initial beam in $P_2$, and close to the 12\,$\mu$m run in $L_g$ and $P_3/P_1$ (7.9\,cm, 2.3\,\%). Figure~\ref{fig:fel}(a) shows that the agreement holds along the whole undulator, not only at saturation. The comparison is not like for like: the codes differ in space charge (and hence in beam size), in the shot noise (Puffin runs with a quiet beam; the early second harmonic of GEM-PIC-RZ, before $z \approx 0.35$\,m, is spontaneous emission from the shot noise) and in the field diagnostics. It is an independent check of the level and the transverse structure of the second harmonic, and in that sense the result of GEM-PIC-RZ is confirmed to within the effect of the beam size; that the second harmonic depends so strongly on the beam size is in line with the analyses of Refs.~\cite{Biedron2002b,Geloni2005}. In Puffin the natural focusing of the planar undulator was switched off to match the other codes.

\begin{figure*}
\includegraphics[width=\textwidth]{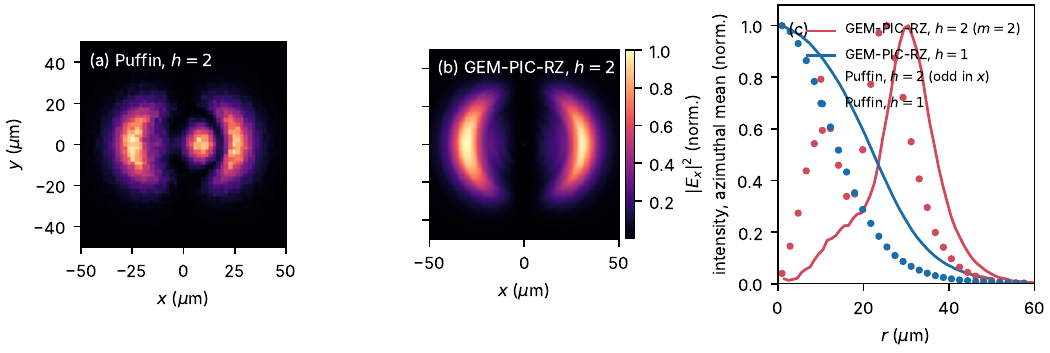}
\caption{Second harmonic near saturation ($z = 0.9$\,m). (a) Puffin, beam of 12\,$\mu$m: $|A_x|^2$ at $2k_r$ in the transverse plane ($x$ is the wiggle plane). (b) GEM-PIC-RZ, beam of 15\,$\mu$m ($\Delta z = \lambda_u/16$, centred orbit, uniform radial grid): $|E_x|^2$ of the radiated second harmonic at $2k_r$ ($m = 2$ and its $m = 0$ part; the co-moving field of the bunching removed, Sec.~\ref{sec:harm}). (c) Azimuthal mean of the intensity against $r$, normalized: radiated second harmonic from GEM-PIC-RZ ($m = 2$ part of the field demodulated at $2k_r$, $\Delta z = \lambda_u/16$) and from Puffin (part of $A_x$ odd in $x$), and the fundamental of both codes.}
\label{fig:profile}
\end{figure*}

\emph{Transverse structure.} Figure~\ref{fig:profile} shows the second harmonic near saturation in the transverse plane. In Puffin it consists of two lobes along the wiggle plane and is zero on the axis, the $E_x \propto x$ mode; a small spot near the centre of the beam, even in $x$, is the field driven by the transverse current at $2k_r$ (the $m = 1$ part discussed above). Azimuthally averaged, the radiated mode is a ring: at $r = 25\,\mu$m in Puffin with a 12\,$\mu$m beam, at $30\,\mu$m in GEM-PIC-RZ, where the beam has grown to 15\,$\mu$m by $z = 0.9$\,m; the fundamental is wider in GEM-PIC-RZ for the same reason. The ring lies well outside the beam, as expected for a mode driven by the transverse gradient of the bunched beam. The map of GEM-PIC-RZ [Fig.~\ref{fig:profile}(b)] shows the same two lobes, zero on the axis; its $y$-polarized part is 1.6\,\% of the $x$-polarized one. A stretched radial grid and an offset orbit both spoil this picture. In earlier runs, with the orbit offset by $x_w$ by the one-period entrance ramp and the grid stretched from $\Delta r = 1$ to 20\,$\mu$m beyond $r = 40\,\mu$m, the $m = 2$ amplitude inside the ring carried a fine radial modulation (rms 20\,\% of the local intensity) with $k_\perp \approx 1.25\,\mu$m$^{-1}$. This is the cone $k_\perp = (2\cdot 2k_r k_u)^{1/2}$ on which radiation at $2\omega_1$ satisfies the resonance with one undulator wavenumber of slippage, i.e.\ third-harmonic emission at $\theta \approx 1.0$\,mrad red-shifted to $2\omega_1$. The offset orbit breaks the rotation-by-$\pi$ symmetry behind the parity rule and couples this odd-harmonic emission into the even modes, and the stretched grid, whose cells beyond 40\,$\mu$m cannot carry $k_\perp \approx 1.25\,\mu$m$^{-1}$, reflected it back towards the axis. With the orbit centred and a uniform grid (the reference run) the cone is gone (its spectral band holds 1.5--1.9\,\% of the $m = 2$ power instead of 16--18\,\%), the $y$-polarized part of the radiated second harmonic drops from 1.6 to 0.2\,\%, the $m = 1$ share from 6.6 to 0.3\,\%, and the ring and the powers change by less than 5\,\%. A radial grid must therefore not stretch where radiation with $k_\perp\Delta r \gtrsim 1$ still has to travel outwards.

\begin{table}
\caption{Seeded planar FEL at saturation.}
\label{tab:fel}
\begin{ruledtabular}
\begin{tabular}{lccc}
 & GENESIS 4 & Puffin & GEM-PIC-RZ \\
\hline
saturation point & 0.84\,m & 0.85\,m & 0.94\,m \\
$P_1$ & 9.8\,GW & 8.3\,GW & 10.6\,GW \\
power gain length & 7.0\,cm & 6.9\,cm & 7.9\,cm \\
$P_2$ & 0 (not modelled) & 11\,MW & 12\,MW \\
$P_3$ & 0.27\,GW & 0.22\,GW & 0.24\,GW \\
$b_1$, $b_2$, $b_3$ & 0.66, 0.25, 0.13 & 0.65, 0.21, 0.11 & 0.63, 0.20, 0.14 \\
space charge & no & no & yes \\
\end{tabular}
\end{ruledtabular}
\end{table}

\emph{Step size in planar undulators.} The mean-field push of the beam (Sec.~\ref{sec:model}) is what makes $\Delta z = \lambda_u/16$ usable here. Pushed in the fields of $s^0$ alone, the 4\,kA beam blew up transversely from the tail forward, to a normalized emittance of 2.8\,$\mu$m and a size of 21\,$\mu$m after 25 periods, and the gain length came out 1.8 times too long; $\lambda_u/32$ was needed (0.04\,$\mu$m). With the mean-field push the emittance after 25 periods is 0.04\,$\mu$m at $\lambda_u/16$ and 0.08\,$\mu$m at $\lambda_u/8$, and the full run at $\lambda_u/16$ agrees with the reference run at $\lambda_u/32$: gain length 7.99 vs 7.89\,cm, $P_1$ at saturation 10.7 vs 10.6\,GW, $P_3/P_1$ 2.27\,\% in both, $P_2$ 11.8 vs 11.6\,MW. On the stretched radial grid of the earlier runs (122 nodes, 1\,$\mu$m up to 40\,$\mu$m, then growing to 30\,$\mu$m at the wall) the same $\lambda_u/16$ run took 1.5\,h on two cores (58\,min without the refinement near the axis, which changes $P_2$ by 4\,\%), with $P_1$, $L_g$ and $P_3/P_1$ unchanged within 1\,\% and $P_2$ within 2\,\%. The second harmonic is the most sensitive quantity: on the stretched grid, the run at $\lambda_u/32$ with the first-order push gave 9.6\,MW, 13\,\% below the mean-field push. The helical undulator is far less sensitive (0.33\,$\mu$m after 25 periods at $\lambda_u/16$ with the first-order push).

\section{When does period averaging fail?}\label{sec:averaging}

\emph{Why.} A natural expectation is that unaveraged codes are needed when the gain per period is large, $\rho \gtrsim 0.01$, as in compact FELs driven by plasma accelerators~\cite{Wang2021,Labat2023,Pompili2022}. We tested this in one dimension.

\emph{Model.} The steady-state 1D FEL equations of a planar undulator were integrated (RK4, 96 steps per period, 256 particles per bucket) in two forms: the averaged equations with the coupling $K[JJ]/2$ and the mean longitudinal velocity, and the unaveraged equations with $v_x = cK\cos(k_u z)/\gamma$ and the figure-eight oscillation of $v_z$, the field remaining a single-frequency slowly varying amplitude. The test thus addresses only the period averaging of the electron motion, not the radiation model, broadband effects or short bunches.

\begin{figure}
\includegraphics[width=\columnwidth]{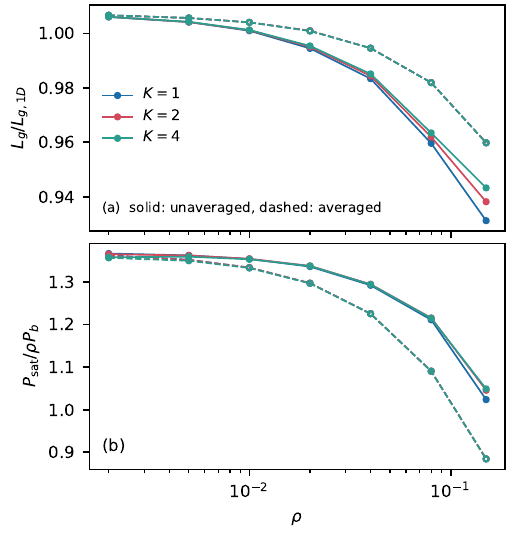}
\caption{1D steady-state planar FEL with averaged (dashed) and unaveraged (solid) electron motion: (a) power gain length and (b) saturation power, against $\rho$.}
\label{fig:averaging}
\end{figure}

\emph{Results.} Figure~\ref{fig:averaging}: both reproduce $L_{g,1D} = \lambda_u/(4\pi\sqrt3\rho)$ and $P_\mathrm{sat} = 1.36\rho P_b$ for $\rho \to 0$. Up to $\rho = 0.15$, where the field grows $e$-fold in a third of a period, the gain lengths differ by less than 3\,\%, independently of $K$; the saturation power of the averaged model is low by about $1.2\rho$ (5\,\% at $\rho = 0.04$). Period averaging is therefore not the main limitation of averaged codes even at large $\rho$; what they omit are the even harmonics and the full space-charge and radiation fields. These are steady-state results; in short bunches and superradiant spikes, where the slippage over a gain length is comparable to the bunch, the slice model of averaged codes may matter more, and this is left for future work.

\section{Discussion and outlook}\label{sec:disc}

\emph{Cost.} We quote the cost of the planar run of Sec.~\ref{sec:harm} on the stretched radial grid, which gives the powers and gain lengths within 5\,\% of the uniform-grid reference (Sec.~\ref{sec:harm}); the uniform grid is needed only for the fine structure of the second harmonic and could be made much cheaper with an absorber in place of the outer cells. That run has 18\,372 slices of $\lambda_r/32$ over a window of 6.1\,$\mu$m, 122 radial nodes, three azimuthal modes and 640\,000 macroparticles; at $\Delta z = \lambda_u/16$ it took 58\,min on two CPU cores and 5.6\,GB of memory. The beam push, done twice per step, costs about as much as the field solve. A steady-state GENESIS run of one slice takes 25\,s with the $401^2$ grid of the comparison and 10\,s with $201^2$, which is converged to 0.3\,\% in $P_1$, $P_3$ and $L_g$. A time-dependent GENESIS run over the same bunch (302 slices of $\lambda_r$ covering the bunch and its slippage, fields at $h = 1, 2, 3$, 8192 particles per slice) took 67\,min on two cores with the $401^2$ grid and 28\,min with the converged $201^2$ grid. GEM-PIC-RZ at $\lambda_u/16$ took 58\,min, or 91\,min with the refinement near the axis: two to three times the averaged code. The window of GEM-PIC-RZ also holds the bunch edges and the space between the bunch and the window ends that its own field needs, and it computes all harmonics and the full space charge at once. For hard X-rays the slices of $\lambda_r/32$ become the limit: a 10\,$\mu$m bunch at 1\,\AA\ needs $3\times10^6$ slices. The march along $\zeta$ is pipelined over MPI ranks, and the particles and field updates within a slice run in parallel on a GPU~\cite{GEMPICRZ_paper}, but such runs are a task for a cluster, not for two cores.

\emph{Limits.} The model drops the slow derivative of the backward field $L$, which is computed from its $\zeta$-equation at each level; radiation travelling backwards, as in an oscillator, is not described. A harmonic $h$ needs the azimuthal modes up to $m = h$ for its off-axis part, the third harmonic $M = 3$. The space charge cannot be switched off, which is physics but makes comparisons with codes that omit it indirect, as Sec.~\ref{sec:harm} shows. The step must resolve the undulator period, $\lambda_u/16$ with the mean-field push of the beam.

\emph{Outlook.} The same code simulates the plasma accelerator, the transport and the undulator. A chained run, with the witness bunch of a laser wakefield stage written at the plasma exit and re-imported, focused by a plasma lens or solenoid and decompressed in a chicane before the undulator, is the natural next step, as are short bunches and superradiance, where the slippage dominates, and harmonics carrying orbital angular momentum in helical undulators, which live in the modes $m \ge 2$ that a mode code computes directly.

\section{Conclusions}\label{sec:concl}

The Galilean electromagnetic PIC method simulates free-electron lasers without period averaging: the electrons follow their orbits within the undulator period, the forward radiation is resolved with its carrier, and the space charge is computed from the same field equations, without a Courant limit on the step; backward radiation, as in an oscillator, is outside the model. In the quasi-cylindrical code GEM-PIC-RZ the harmonics of a planar undulator separate by azimuthal mode, odd harmonics in odd modes and even harmonics in even modes. The code reproduces the coherent spontaneous emission of a planar undulator in each of the modes $m = 0$--3 to within 2\,\% of the radiation integral and seeded amplifiers to within a few per cent of GENESIS with the same physics. For a seeded planar FEL at 10.6\,nm it predicts a second harmonic of 12\,MW at saturation, $1.1\times10^{-3}$ of the fundamental, which an averaged code does not model; the unaveraged code Puffin, which has no space charge and therefore a different beam size, gives a comparable level, 9 to 11\,MW depending on the beam size, and the same transverse structure. Period averaging itself is accurate for the fundamental up to large FEL parameters, $\rho = 0.15$ in one dimension. What averaged codes leave out by construction are the even harmonics and the full fields of the beam, and these are what an unaveraged, carrier-resolved code adds.

\section*{Data availability}
The input decks of all GEM-PIC-RZ and GENESIS runs in this paper are part of the code repository~\cite{GEMPICRZ} (\texttt{benchmarks/fel\_paper}).

\begin{acknowledgments}
This work was supported by DFG and BMBF. The author gratefully acknowledges the Gauss Centre for Supercomputing e.V. (gauss-centre.eu) for funding this project (lpqed) by providing computing time through the John von Neumann Institute for Computing (NIC) on the GCS Supercomputer JUPITER at J\"ulich Supercomputing Centre (JSC). Parts of the code and of the analysis were developed with the assistance of the AI model Claude~\cite{Claude}.
\end{acknowledgments}


\begin{thebibliography}{99}
\bibitem{Pellegrini2016} C. Pellegrini, A. Marinelli, and S. Reiche, The physics of x-ray free-electron lasers, Rev. Mod. Phys. \textbf{88}, 015006 (2016).
\bibitem{KimHuangLindberg} K.-J. Kim, Z. Huang, and R. Lindberg, \textit{Synchrotron Radiation and Free-Electron Lasers} (Cambridge University Press, Cambridge, 2017).
\bibitem{Reiche1999} S. Reiche, GENESIS 1.3: a fully 3D time-dependent FEL simulation code, Nucl. Instrum. Methods Phys. Res., Sect. A \textbf{429}, 243 (1999).
\bibitem{Genesis4} S. Reiche, GENESIS 1.3 version 4, \url{https://github.com/svenreiche/Genesis-1.3-Version4}; release 4.6.15 was used.
\bibitem{Bonifacio1984} R. Bonifacio, C. Pellegrini, and L. M. Narducci, Collective instabilities and high-gain regime in a free electron laser, Opt. Commun. \textbf{50}, 373 (1984).
\bibitem{Huang2000} Z. Huang and K.-J. Kim, Three-dimensional analysis of harmonic generation in high-gain free-electron lasers, Phys. Rev. E \textbf{62}, 7295 (2000).
\bibitem{Freund2000} H. P. Freund, S. G. Biedron, and S. V. Milton, Nonlinear harmonic generation and proposed experimental verification in SASE FELs, Nucl. Instrum. Methods Phys. Res., Sect. A \textbf{445}, 53 (2000).
\bibitem{Geloni2005} G. Geloni, E. Saldin, E. Schneidmiller, and M. Yurkov, Exact solution for the second harmonic generation in XFELs, DESY 05-137 (2005), arXiv:physics/0508046.
\bibitem{Tremaine2002} A. Tremaine \textit{et al.}, Experimental characterization of nonlinear harmonic radiation from a visible self-amplified spontaneous emission free-electron laser at saturation, Phys. Rev. Lett. \textbf{88}, 204801 (2002).
\bibitem{Biedron2002} S. G. Biedron \textit{et al.}, Measurements of nonlinear harmonic generation at the Advanced Photon Source's SASE FEL, Nucl. Instrum. Methods Phys. Res., Sect. A \textbf{483}, 94 (2002).
\bibitem{Biedron2002b} S. G. Biedron \textit{et al.}, Impact of electron beam quality on nonlinear harmonic generation in high-gain free-electron lasers, Phys. Rev. ST Accel. Beams \textbf{5}, 030701 (2002).
\bibitem{Campbell2012} L. T. Campbell and B. W. J. McNeil, Puffin: a three dimensional, unaveraged free electron laser simulation code, Phys. Plasmas \textbf{19}, 093119 (2012); version 2.1.0a, \url{https://github.com/UKFELs/Puffin}, was used.
\bibitem{Fallahi2018} A. Fallahi, A. Yahaghi, and F. X. K\"artner, MITHRA 1.0: a full-wave simulation tool for free electron lasers, Comput. Phys. Commun. \textbf{228}, 192 (2018).
\bibitem{FawleyVay2009} W. M. Fawley and J.-L. Vay, Use of the Lorentz-boosted frame transformation to simulate free-electron laser amplifier physics, AIP Conf. Proc. \textbf{1086}, 346 (2009).
\bibitem{Fallahi2020} A. Fallahi, MITHRA 2.0: a full-wave simulation tool for free electron lasers, arXiv:2009.13645 (2020).
\bibitem{Vay2007} J.-L. Vay, Noninvariance of space- and time-scale ranges under a Lorentz transformation and the implications for the study of relativistic interactions, Phys. Rev. Lett. \textbf{98}, 130405 (2007).
\bibitem{GEMPIC} A. Pukhov, N. Elkina, and T. Wilson, Galilean electromagnetic particle-in-cell code, \href{https://doi.org/10.1016/j.jcp.2026.115331}{J. Comput. Phys. \textbf{567}, 115331 (2026)}; arXiv:2509.16793.
\bibitem{GEMPICRZ_paper} A. Pukhov, GEM-PIC-RZ: a Galilean electromagnetic particle-in-cell code in quasi-cylindrical geometry with nonuniform grids, arXiv:2610.31622 (2026).
\bibitem{GEMPICRZ} A. Pukhov, GEM-PIC-RZ, version 1.0.0, \href{https://doi.org/10.5281/zenodo.23600179}{doi:10.5281/zenodo.23600179} (2026); \url{https://github.com/AlexMPukhov/GEM-PIC-RZ}, benchmarks in \texttt{benchmarks/}.
\bibitem{Vay2008} J.-L. Vay, Simulation of beams or plasmas crossing at relativistic velocity, Phys. Plasmas \textbf{15}, 056701 (2008).
\bibitem{Fawley2002} W. M. Fawley, Algorithm for loading shot noise microbunching in multidimensional, free-electron laser simulation codes, Phys. Rev. ST Accel. Beams \textbf{5}, 070701 (2002).
\bibitem{Jackson} J. D. Jackson, \textit{Classical Electrodynamics}, 3rd ed. (Wiley, New York, 1999).
\bibitem{Xie1995} M. Xie, Design optimization for an X-ray free electron laser driven by SLAC linac, in \textit{Proc. PAC'95} (IEEE, 1995), p.~183.
\bibitem{Wang2021} W. Wang \textit{et al.}, Free-electron lasing at 27 nanometres based on a laser wakefield accelerator, Nature \textbf{595}, 516 (2021).
\bibitem{Labat2023} M. Labat \textit{et al.}, Seeded free-electron laser driven by a compact laser plasma accelerator, Nat. Photon. \textbf{17}, 150 (2023).
\bibitem{Pompili2022} R. Pompili \textit{et al.}, Free-electron lasing with compact beam-driven plasma wakefield accelerator, Nature \textbf{605}, 659 (2022).
\bibitem{Claude} Anthropic, Claude, large language model (model \texttt{claude-opus-5-5}), \url{https://www.anthropic.com/claude}, used in October 2026.
\end{thebibliography}
\end{document}